\newcommand\rxte{{\it RXTE}}
\newcommand\xmm{{\it XMM-Newton}}
\newcommand\ftool{{\it FTOOL}}

\newcommand\kev{{\rm~keV}}

\newcommand\flux{\ifmmode {\rm~erg cm}$^{-2}$\ ; {\rm s}$^{-1}$ \else ~erg cm$^{-2}$ s$^{-1}$\fi}
\newcommand\kms{\ifmmode {\rm~km\ s}$^{-1}$ \else ~km s$^{-1}$\fi}
\newcommand\Hunit{\ifmmode {\rm~km\ s}$^{-1}$\ {\rm Mpc}$^{-1}$
        \else ~km s$^{-1}$ Mpc$^{-1}$\fi}
\newcommand\ctssec{\ifmmode {\rm~count\ s}$^{-1}$ \else ~count s$^{-1}$\fi}
\newcommand\ergsec{\ifmmode {\rm~erg\ s}$^{-1}$ \else
        ~erg s$^{-1}$\fi}
\newcommand\funit{\ifmmode {\rm~erg\ s}$^{-1}$\ ; {\rm cm}$^{-2}$ \else
        ~ergs s$^{-1}$ cm$^{-2}$\fi}
\newcommand\phflux{\ifmmode {\rm~photon\ s}$^{-1}$\  ; {\rm cm}$^{-2}$
        \else   ~photon s$^{-1}$ cm$^{-2}$\fi}
\newcommand\efluxA{\ifmmode {\rm~erg\ s}$^{-1}$\ ; {\rm cm}$^{-2}$\ ; {\rm
        \AA}$^{-1}$ \else ~erg s$^{-1}$ cm$^{-2}$ \AA$^{-1}$\fi}
\newcommand\efluxHz{\ifmmode {\rm~erg\ s}$^{-1}$\ ; {\rm cm}$^{-2}$\ ; {\rm
        Hz}$^{-1}$ \else ~erg s$^{-1}$ cm$^{-2}$ Hz$^{-1}$\fi}
\newcommand\cc{\ifmmode {\rm~cm}$^{-3}$ \else cm$^{-3}$\fi}
\newcommand\fwhm{\ifmmode {\rm~FWHM} \else ${\rm~FWHM}$\fi}
\newcommand\msun{\ifmmode M_{\odot} \else $M_{\odot}$\fi}
\newcommand\lsun{\ifmmode L_{\odot} \else $L_{\odot}$\fi}

\newcommand\hbeta{\ifmmode {\rm H}\beta \else H$\beta$\fi}
\newcommand\kalpha{\ifmmode {\rm K}\alpha \else K$\alpha$\fi}
\newcommand\lalpha{\ifmmode {\rm L}\alpha \else L$\alpha$\fi}
\newcommand\nh{\ifmmode N_{\rm H} \else N$_{\rm H}$\fi}
\newcommand\xray{{\rm}X--ray}
\newcommand\hhh{{\rm}1H0707--495}
\newcommand\iras{{\rm}IRAS 13224--3809}
\documentclass[useAMS,usenatbib,10pt,referee]{raa}
\usepackage{times}
\usepackage{url}
\usepackage[colorlinks=true,linkcolor=blue]{hyperref}
\usepackage{float}
\usepackage{natbib}
\usepackage[dvips]{graphicx}
\DeclareGraphicsExtensions{.pdf,.png,.jpg,.mps,.eps,.ps}

\begin{document}
\title{On the reality of broad iron L lines from the  narrow line Seyfert 1 galaxies \hhh{} and \iras{}}
   \volnopage{Vol.0 (200x) No.0, 000--000} 
   \setcounter{page}{1}
   \author{P. K. Pawar
	\inst{1}*
	\and G. C. Dewangan
	\inst{2}
	\and M. K. Patil 
	\inst{1}
	\and R. Misra
	\inst{2}
	\and S. K. Jogadand
	\inst{1}
	}
   \institute{S. R. T. M. University, Nanded -- 431 606 India; {\it pawar.pk123@gmail.com}\\
	\and
		Inter-University center for Astronomy and Astrophysics, Pune -- 411 007\\
	}

\abstract{We performed time resolved spectroscopy of \hhh{} and \iras{} using long \xmm{} observations. These are strongly variable narrow line Seyfert 1 galaxies and show broad features around 1 \kev{} that has been interpreted as relativistically broad Fe \lalpha{} lines. 
Such features are not clearly observed in other AGN despite sometimes having high iron abundance required by the best fitted blurred reflection models. Given the importance of these lines, we explore the possibility if rapid variability of spectral parameters may introduce broad bumps/dips artificially in the time averaged spectrum, which may then be mistaken as broadened lines. We tested this hypothesis by performing time resolved spectroscopy using long ($>$ 100 $ks$) \xmm{} observations and by dividing it into segments with typical exposure of few $ks$. 
We extracted spectra from each such segment and modelled using a two component phenomenological model consisting of a power law to represent hard component and a black body to represent the soft emission. As expected both the sources showed variations in the spectral parameters. Using these variation trends, we simulated model spectra for each segment and then co-added to get a combined simulated spectrum. 
In the simulated spectra, we found no broad features below 1 \kev{} and in particular no deviation near 0.9 \kev{} as seen in the real average spectra.  This implies that the broad Fe \lalpha{} line that is seen in the spectra of these sources is not an artifact of the variation of spectral components and hence providing evidence that the line is indeed genuine.}

   \authorrunning{Pawar et al.}
   \titlerunning{Reality of iron L lines}
   \maketitle
   \date{\today}

\section{Introduction}
Active galactic nuclei (AGN) are thought to be powered by the accretion of matter onto the central super massive black hole (SMBH). The surrounding matter forms an optically thick, geometrically thin disk that radiate mainly in the optical/UV region. The broadband \xray{} spectrum of AGN follows a power law shape and is thought to originate from the Compton up-scattering of low energy disk photons by the relativistic electron cloud present in the hot Comptonizing corona \citep{1976ApJ...204..187S,1985ApJ...289..514Z,1980A&A....86..121S,1991ApJ...380L..51H}. 
The geometry and origin of this corona is, however, still unclear. In addition to this primary power law continuum several other features are also apparent, which include reflection hump in the energy range 10--50 \kev{}, broad and skewed Fe \kalpha{} fluorescent line around 6.4 \kev{} and soft excess emission below 1 \kev{}. Origin of the reflection hump and the fluorescent Fe \kalpha{} line is generally attributed to the reflection of the power law photons by the relatively cold accretion disk \citep{1991MNRAS.249..352G}. 
First clear evidence for the presence of the extremely broad, skewed iron line came from the {\it ASCA} long observation of MCG--6--30--15 \citep{1995Natur.375..659T}. Similar line profiles were later noticed in several other AGN (for a review, see \citealt{2007ARA&A..45..441M} and references therein). Detection of the broad Fe \kalpha{} is important as it carries signature of the inner accretion disk close to the SMBH e.g. the line energy tells us the ionization state of the disk, the inner radius of accretion disk can be inferred from the redward wing of the line which extends sometimes even down to $\sim$ 2 \kev{}. The line also provides an unique tool to test and verify the GR theory \citep{2000PASP..112.1145F}.

As mentioned earlier, the AGN spectrum consists of features like soft excess emission, reflection hump and broad Fe K lines. These features can be reproduced using the blurred reflection of primary continuum from partially ionized accretion disk \citep{1991MNRAS.249..352G}. 
Several versions of self--consistent reflection models are available e.g. $reflionx$, $relxill$ etc. which provide good approximation to the observed data \citep{2005MNRAS.358..211R,2006MNRAS.365.1067C,2014ApJ...782...76G}. Assuming lamp--post geometry, the recent blurred reflection models describe both the spectral shape and the observed variability of iron lines \citep{2003MNRAS.344L..22M,2004MNRAS.353.1071F,2010MNRAS.401.2419Z,2014MNRAS.443.1723P,2015MNRAS.446..633G}. 
The AGN \xray{} spectra, however, can also be fitted using complex partial--covering absorption model which require the central engine to be obscured by complex absorbing clouds/zones having varying column density, covering fraction and ionization. Here, the observed spectral variability is attributed to the motions of these absorbing clouds \citep{2008A&A...483..437M,2009A&ARv..17...47T,2014PASJ...66..122M,2014ApJ...787...83M,2014MNRAS.441.1817P}.

Both the blurred reflection and partial covering absorption models provide statistically comparable spectral fits and hence must be judged based on the best fit spectral parameter values. One of the key fitting parameter in reflection scenario is the iron abundance.
Observationally, many AGN require over abundance of iron relative to the solar value e.g. Ark 120 \citep{2014MNRAS.439.3016M}; NGC 1365, \citep{2010MNRAS.408..601W}; \hhh{}, \citep{2009Natur.459..540F}. This over abundance of iron must also produce the accompanying Fe \lalpha{} which has not been observed in those AGN spectra. These anomalies cast shadows on the reality of the iron lines and hence on the reflection scenario. Recently, the accompanied Fe \lalpha{} line was detected in two extreme cases of typical narrow line Seyfert 1 galaxies (NLS1) \hhh{} and \iras{} using \xmm{} data by \cite{2009Natur.459..540F} and \cite{2010MNRAS.406.2591P}, respectively.
In fact, \cite{2009Natur.459..540F}, using Fe \lalpha{} line claimed a lag of 30$s$ between the direct power law component (1--4 \kev{}) and the reflection component (0.3--1 \kev{}), as it provided better statistics compared to the \kalpha{} line. However, the AGN spectrum can be complex due to the presence of multi--component absorber and soft \xray{} excess and strong spectral variability may introduce artificial spectral features in the mean spectrum. Artifacts in the AGN spectra introduced due to the spectral modelling have been reported earlier e.g. the soft excess is an artifact of the deficit of emission due to smeared wind absorption while the complex partial-covering absorption model can mimic the broad Fe \kalpha{} line quite well. Both these artifacts are model dependent.
The \xray{} spectrum of NLS1 is variable and can vary within few $ks$ \citep{1996A&A...305...53B}. In the case of \xmm{} observations, where typical exposure time is $>$ 100 $ks$, it is very likely that both the flux and the spectral shape may change significantly. These changes in the flux and/or spectral parameters may introduce the broad bumps/dips in the time averaged spectrum. In this paper, using a two component phenomenological model and using time resolved spectroscopy, we try to comment on the reality of Fe \lalpha{} line observed in these two AGN. The observational details and data reduction techniques are summarized in section 2. The spectral analysis and the results are presented in section 3. The discussion and conclusion is given in section 4.

\section{Observations and data reduction}
\hhh{} has been observed with \xmm{} \citep{2001A&A...365L...1J} on various occasions, particularly continuous observations were carried out during 29 January - 6 February, 2008 and 13 - 19 September, 2010. We studied all four long observations from 2008 during which the European Photon Imaging Camera--pn (EPIC--pn; \citealt{2001A&A...365L..18S}) was operated in the large window imaging mode with medium filter while optical monitor (OM; \citealt{2001A&A...365L..36M}) was operated with UVW1 filter in the fast imaging mode and $\sim$ 300 exposures were taken during the monitoring. 
The second source, \iras{}, was observed with \xmm{} four times during 2011 July, 19-29 for more than 500 $ks$ and once on 2002 January 19 for $\sim$ 64 $ks$. We used four 2011 observations for the current study in which the optical monitor was operated in the fast imaging mode with UVM2 filter having $\sim$ 120 exposures. The EPIC--pn camera was operated in large window imaging mode. We performed time resolved spectroscopy using these observations and found consistency in the results derived from them. Hence, in this paper we present results of \hhh{} and \iras{} obtained from the observation IDs 0511580101 and 0673580201, respectively. The choice of presenting results from these observation IDs is mainly driven by having the maximum net useful X--ray exposure time and total number of OM frames. We acquired the Observation Data Files (ODFs) for both the sources from the \emph{HEASARC}\footnote{http://heasarc.gsfc.nasa.gov/cgi-bin/W3Browse/w3browse.pl} archive. The datasets were reprocessed and filtered using the \emph{XMM-Newton} Science Analysis System SASv13.5\footnote{http://xmm.esac.esa.int/external/xmm\_data\_analysis/} with the latest calibration files and following the SAS ABC guide\footnote{http://xmm\_newton.abc-guide}.

\begin{figure}[H]
\centering
\includegraphics[width=14cm,height=8cm]{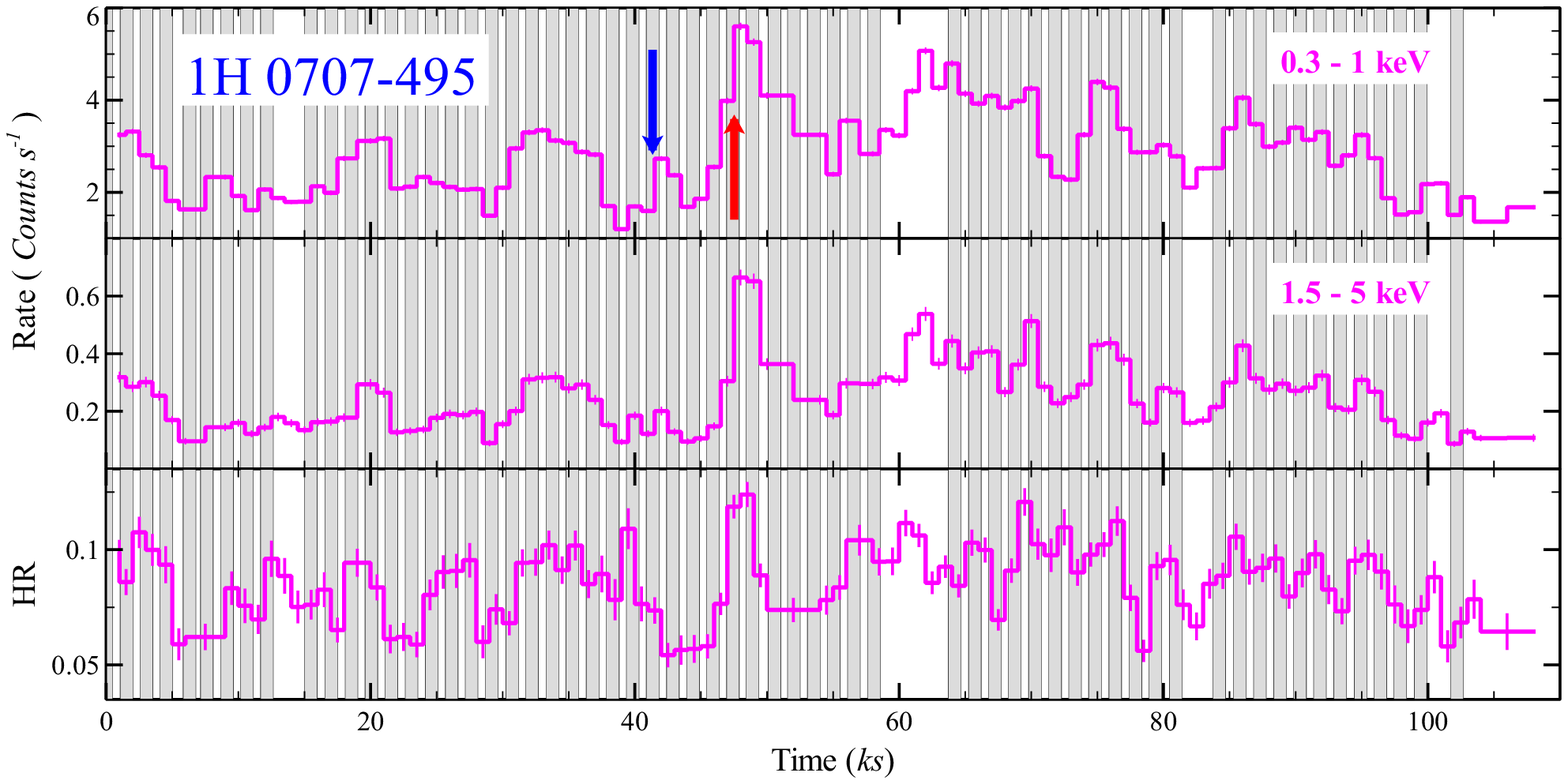}
\includegraphics[width=14cm,height=8cm]{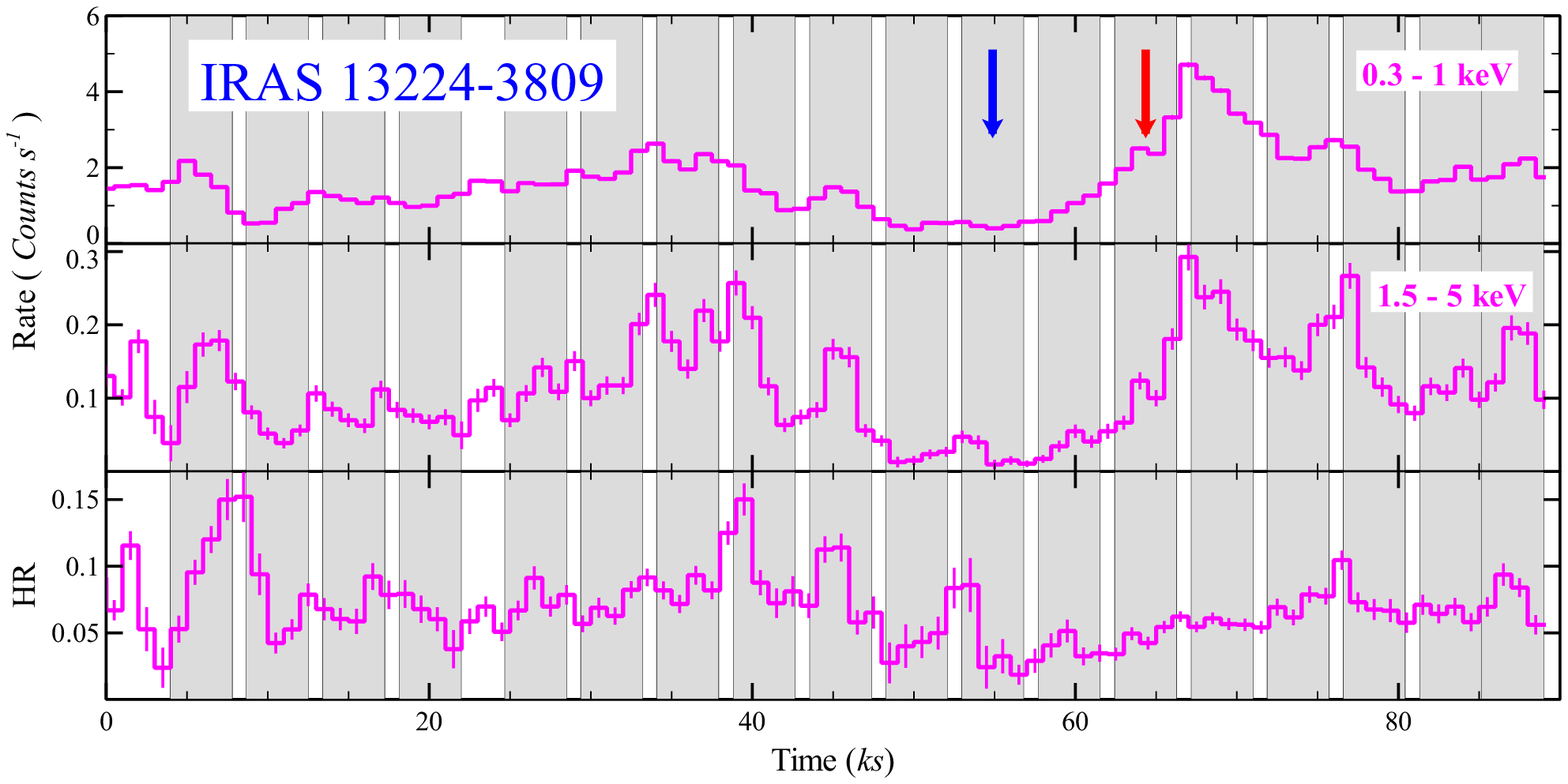}
\caption{Background corrected EPIC--pn lightcurves of \hhh{} ($top$) and \iras{} ($bottom$) depicting variations in 0.3--1 \kev{} soft band ($upper{~\rm }panel$), 1.5--5 \kev{} hard band ($middle{~\rm }panel$), and hardness ratio ($lower{~\rm }panel$). All the lightcurves are binned in 1 $ks$ bins. Gray shade indicates small exposures used for generating segment spectra. The arrows show the exposures used for the generating spectra shown in figure 2 with similar color.}\label{lc_variations}
\end{figure}

\begin{figure*}
\centering
\includegraphics[scale=0.48]{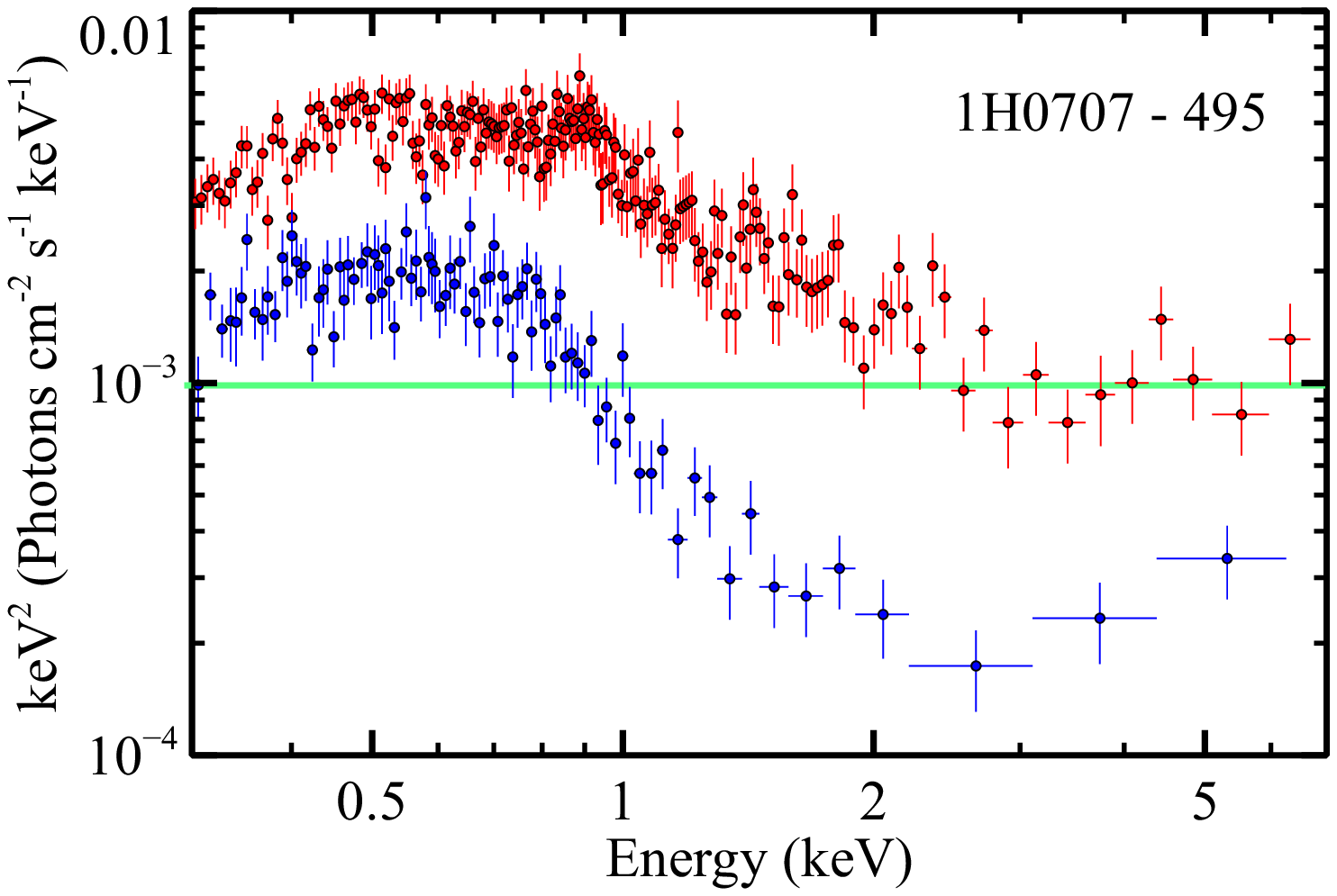}
\includegraphics[scale=0.48]{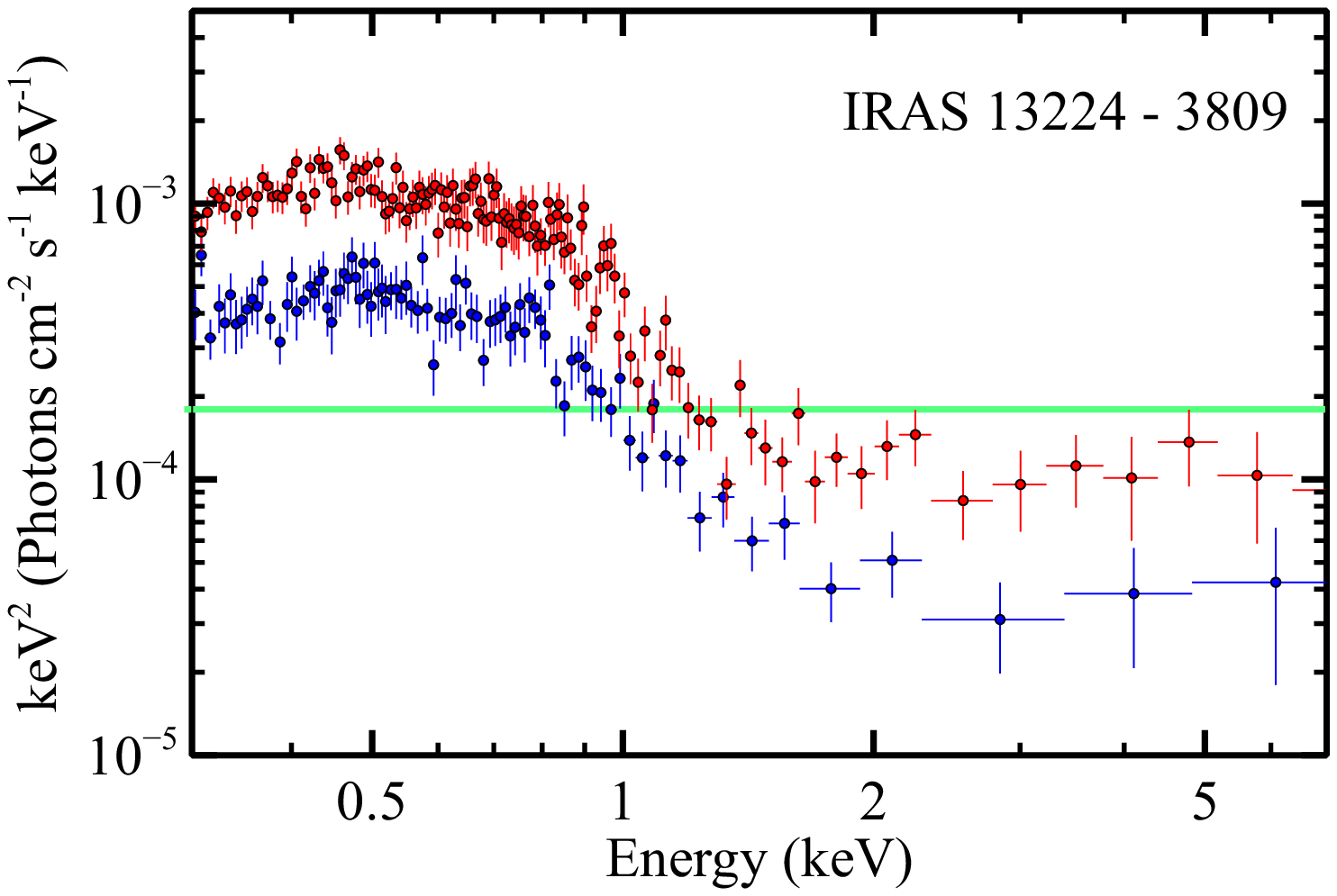}
\caption{The unfolded EPIC--pn spectra derived from two segments of the \hhh{} ($left$) and \iras{} ($right$). These spectra were unfolded with a Power Law model having $\Gamma$ fixed at 2 to demonstrate the spectral variation within the observations for both AGN.}\label{spec_examples}
\end{figure*}

To produce calibrated \xray{} images, we reprocessed both the dataset using the \emph{epproc} pipeline script. \emph{epproc} operates in such a way that it rejects invalid events, determines good time interval and assigns various flags to each event. It also applies corrections due to the charge transfer inefficiency (CTI) and gain variations. We considered good quality single and double events by setting \emph{FLAG} = 0 and \emph{PATTERN} $\leq$ 4. 
By generating background light curves above 10 \kev{}, each dataset was examined for the flaring particle background. The time where the data was severely affected by such background flaring was filtered appropriately. The observations were not affected by pile--up and thus no correction was applied.
We generated the background corrected source lightcurves in the soft (0.3--1 \kev) and hard (1.5--5 \kev) bands which are shown in fig.~\ref{lc_variations}. These lightcurves show significant variability with the fractional variability amplitude of 33 and 56 per cent for \hhh{} and 52 and 57 per cent that for \iras{} in soft and hard bands, respectively. $Lower~panels$ of fig.~\ref{lc_variations} show the variations in the hardness ratio between the two \xray{} bands. Variations in the hardness ratio naturally demonstrates the spectral variations of the source. Due to these short--term variations in the hardness ratio we divided both the observations into smaller segments using the time stamps of the simultaneous OM frames. Typical exposure of each \xray{} segment was $\sim$ 1.2 $ks$ for \hhh{} and 4 $ks$ for \iras{}. From each such segment the source spectrum was accumulated from a circular region of radius 35'' - 40'' centered on the source, whereas the background was extracted from the source free region on the same chip. In fig.~\ref{spec_examples} we plotted two such segment spectra of each source (shown by arrows with similar colors in fig.~\ref{lc_variations}) to demonstrate the spectral variations. The spectra were unfolded by fitting a $power~law$ model with $\Gamma$ fixed at 2. This again shows the spectral variations between the segments.
The EPIC--pn response files i.e. the redistribution matrix files (RMF) and the effective area files (ARF) were generated using the SAS tasks \emph{rmfgen} and \emph{arfgen}, respectively. The resulting spectra were then grouped using \ftool{} task \emph{grppha} to a minimum of 20 counts per spectral bin so that the $\chi ^2$ minimization technique can be employed. All the errors on the best-fit parameters quoted here corresponds to the 1--$\sigma$ range.

\begin{figure*}
\centering
\includegraphics[scale=1]{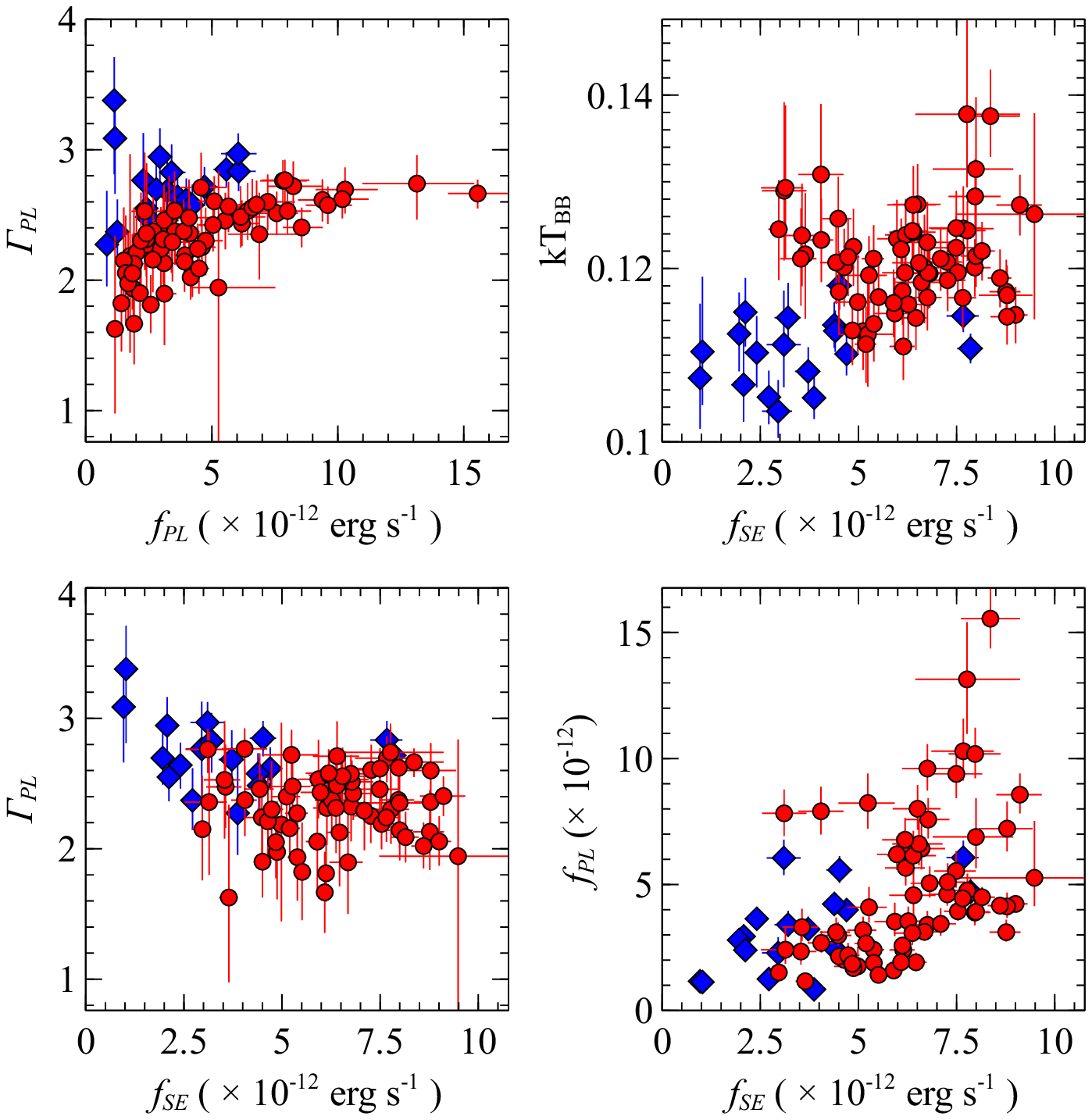}
\caption{Short--term variation of spectral parameters derived from 100 $ks$ long observations of \hhh{} (red box) and \iras{} (blue diamond). The component fluxes, $f_{SE}$ and $f_{PL}$, are derived from 0.3--1 \kev{} and 1.5--3 \kev{} bands, respectively.}\label{spec_variations}
\end{figure*}

\section{Spectral Analysis}\label{analysis_method}
The \xray{} spectrum of AGN, particularly of NLS1, is considered to be a two component spectra representing hard/primary emission and soft emission.
To model the spectra generated from each \xray{} segment, we used two component phenomenological model ($powerlaw + bbody$) corrected for Galactic absorption, with the power law (PL) component to constrain the continuum or hard emission (PL), while the black body component to represent the soft emission (SE). We fitted the entire 0.3--10 \kev{} band with the $wabs*(powerlaw + bbody)$ model. The component fluxes were calculated using $cflux$ model. We noted the best fit values of photon index, disk temperature and the component fluxes ($f_{SE}$ in 0.3--1 \kev{} and $f_{PL}$ in 1.5--3 \kev{} bands). Due to poor signal above 6 \kev{} we could not constrain the Fe \kalpha{} line in both the objects therefore we fitted spectra without line component.
The detailed results of this time resolved spectroscopy (TRS) will be presented in a separate paper (Pawar et al. 2016), which will discuss both the spectral variability as well as the \xray{}/Optical correlation. For the sake of brevity, we describe here the quality of fits and the observed variations. Most of the spectra were well fitted using two component model resulting in the reduced $\chi^2$ between 0.8--1.2. Any fit whose reduced $\chi^2$ lies outside of these values were termed bad fit and were not considered in the further analysis. This resulted in total 68 and 19 spectra from single observations for \hhh{} and \iras{}, respectively. From this study, it has been observed that both the PL and SE components are variable. Fig.~\ref{spec_variations} shows the intra--observation variability of spectral parameters of both the components for \hhh{} (red circles) and \iras{} (blue diamonds). From the $\Gamma_{PL}$ - $f_{PL}$ plot, both the objects follow the softer when brighter nature. This nature is already studied by \cite{2009MNRAS.399.1597S} using \rxte{} data of a sample of 10 Seyfert galaxies and found that except for NGC 5548 all other AGN showed softer when brighter behaviour. However, using 24 galaxies from Palomer sample observed by swift, \cite{2016MNRAS.459.3963C} showed that the high luminosity AGN appeared softer when brighter while low luminosity AGN showed harder when brighter trend. Both the objects studied here are NLS1 with high accretion rate indicating that they follow the similar spectral properties that of high luminosity end AGN. Variability of these spectral parameters on shorter timescale may introduce artifacts in the time averaged spectrum over sufficiently longer time. Such averaging may lead to the bumps and dips in the time averaged spectrum causing artificial broad line like features. 
To check this possibility, we simulated fake data using the best fit theoretical model of each segment spectra using the \emph{XSPEC} task \emph{fakeit}.  Later on, the resultant simulated spectra were co-added together using the \emph{FTOOL} \emph{addascaspec} to get the combined simulated spectrum. The response and background files were also merged together. 
The resultant combined simulated spectra of both the objects were used for the further spectral analysis. Similarly we also derived the time averaged spectra from 35\arcsec{} region centered on the source position as well as the background spectra from a region devoid of source contamination. The resultant spectra were grouped so as to use $\chi^2$ minimization technique.

\begin{figure*}
\begin{center}
\includegraphics[scale=0.28,angle=90]{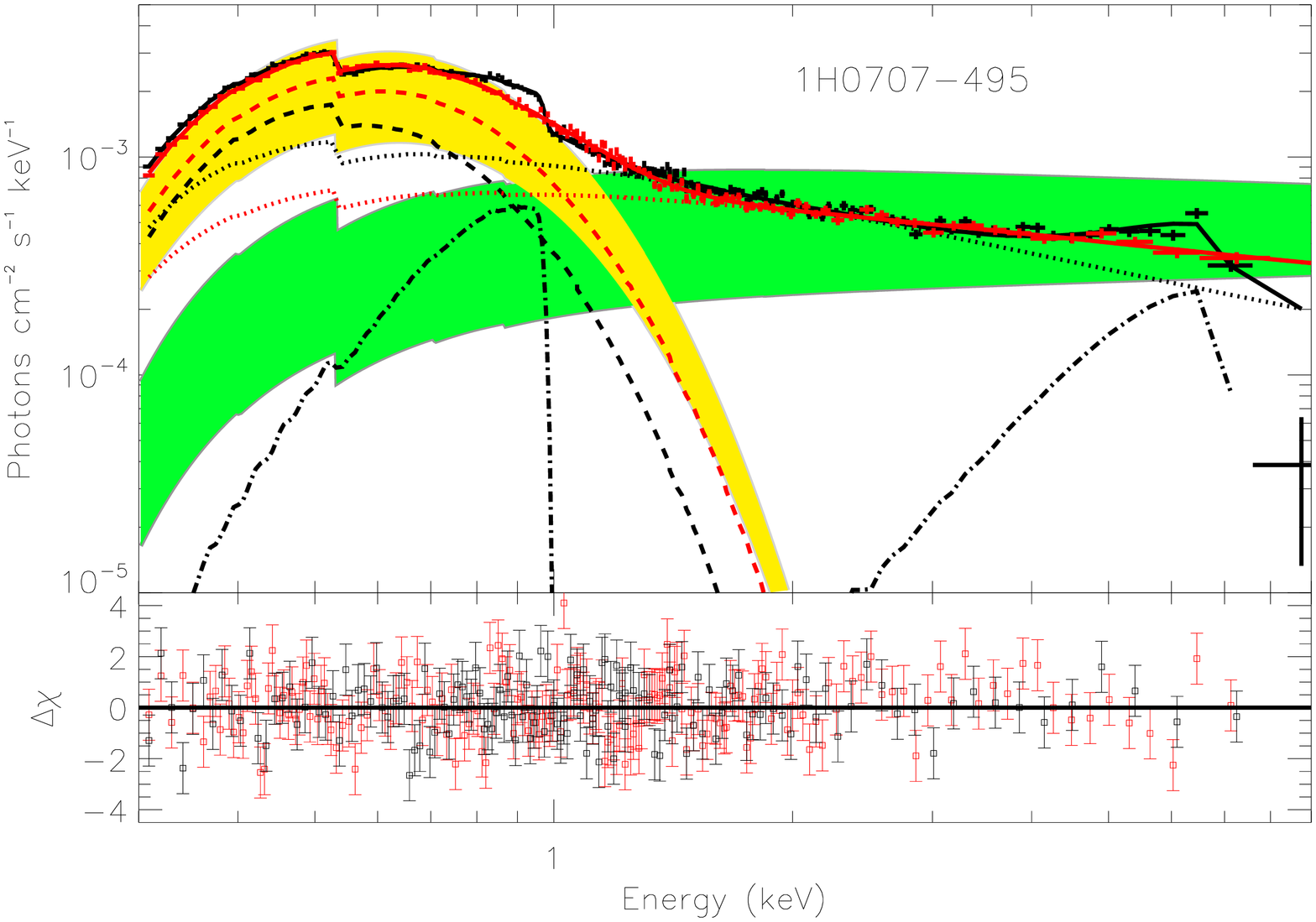}\hspace{0.6cm}
\includegraphics[scale=0.28,angle=90]{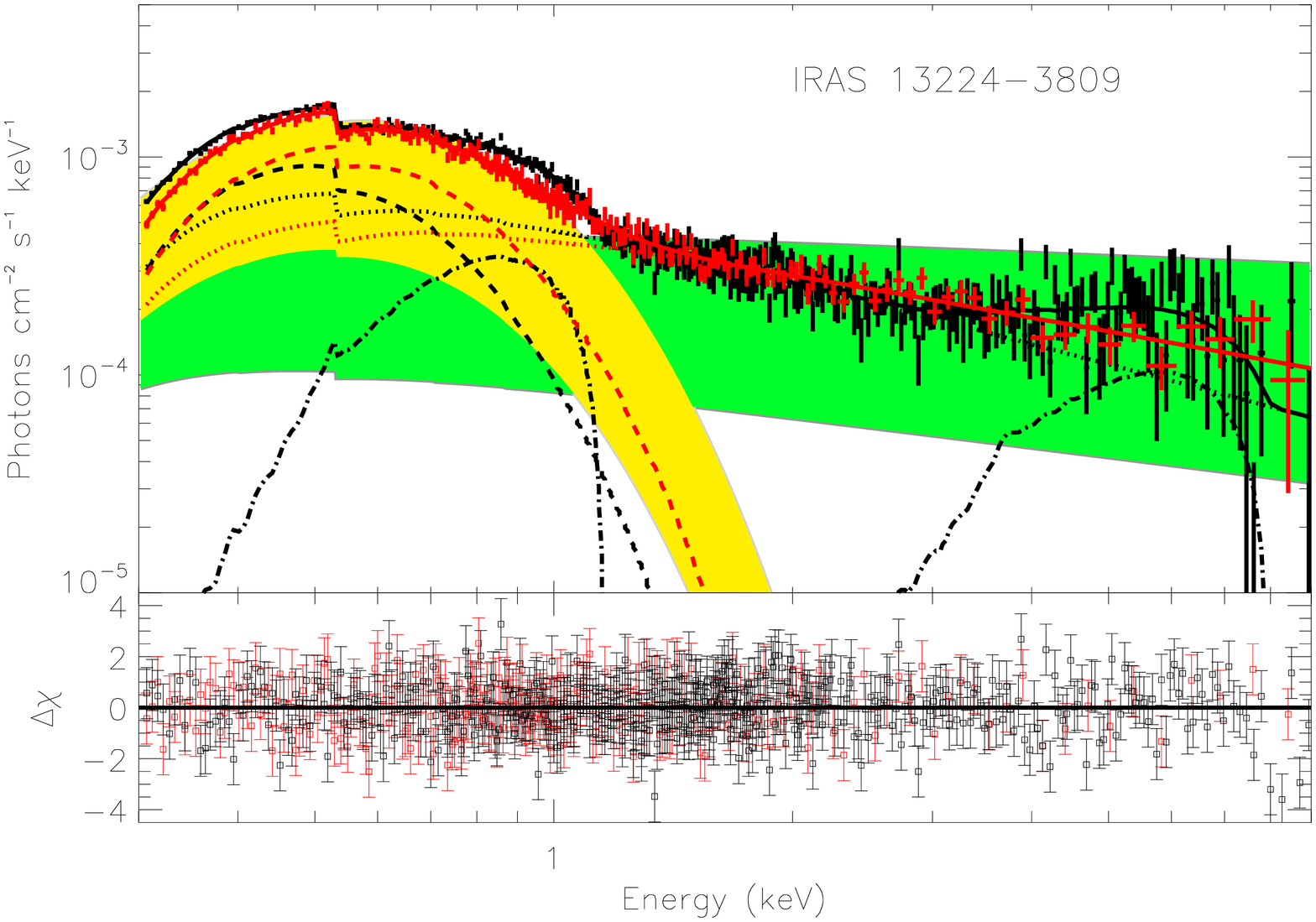}
\caption{The unfolded time averaged spectrum (in black) and simulated spectrum (in red) of \hhh{} (\emph{top}) and \iras{} (\emph{bottom}) were fitted with absorbed power law (dotted lines) and blackbody (dashed lines) components. Two $Laor$ lines (dash-dotted line) are used to model the broad Fe lines. $\chi$ variation is shown in the bottom panel of each figure. The solid thick lines are the best fit model. The green and yellow shaded region represents the area between the lowest and highest flux values of PL and BB components, respectively. The figure for \hhh{} is rebinned for clarity.}\label{felline_real}
\end{center}
\end{figure*}

We applied our phenomenological model to the simulated spectra and average spectra simultaneously. The simulated spectra of both the objects provided excellent fit with no apparent deviations to the model, however, fit to the average spectra was very poor and showed large deviations. To account for these deviations we added two $Laor$ components. Assuming that the origin of both the lines is same, we tied the emissivity index ($\beta$), inner radius (R$_{in}$), outer radius (R$_{out}$) and inclination of both the $Laor$ components to each other. 
The spectral results for both the objects are shown in Table~\ref{spec_res}. The difference in the best fit PL and BB parameters for simulated and time averaged spectra as seen in Table~\ref{spec_res} is probably due to the fact that the good time interval (GTI) of simulated spectra is a subset of GTI of average spectra.
This extra time in the average spectrum might have changed shape of the spectrum. Figure~\ref{felline_real} represent our best fit spectral results. For each of the source, we plotted the data along with the best fit model and the variation of $\chi$ for both the simulated spectra (red) and the average spectra (black). The individual model components power law (dotted lines), blackbody (dashed lines)  and two $laor$ (dot-dash ) lines are also shown. From the figure it is clear that the simulated spectrum does not require the $laor$ line at $\sim$ 1 keV. The green and yellow shaded region indicates the range of variability of power law and black body normalizations, respectively.

\begin{table*}
\begin{center}
\caption[]{Best fit spectral results for \hhh{} and \iras{}. The best fit model for the average spectrum is $wabs*(powerlaw+bbody+laor+laor)$ while for the simulated spectrum is $wabs*(powerlaw+bbody)$. }\label{Tab:publ-works}
\def\arraystretch{1.7}
 \begin{tabular}{c c c c c}
  \hline
	 	&  \multicolumn{2}{c} {\hhh{}} & \multicolumn{2}{c} {\iras{}}\\
Parameters 	&  Average Spectrum     & Simulated Spectrum & Average Spectrum     & Simulated Spectrum\\\hline
Exposure	& 95 $ks$		& 61 $ks$		& 83 $ks$		& 40 $ks$	\\
Net Count Rate	& 3.5 $\pm$ 0.1		& 3.4 $\pm$ 0.1		& 1.9 $\pm$ 0.1		& 1.8 $\pm$ 0.1\\\hline
N$_{H}$ ($\times 10^{20}$ cm $^{-2}$)& 5.9$_{-0.2}^{+0.2}$ 	& 5.1$_{-0.2}^{+0.2}$	& 5.3$_{-0.2}^{+0.2}$   	& 5.4$_{-0.3}^{+0.3}$ \\
$\Gamma$  		& 2.76$_{-0.02}^{+0.03}$	& 2.4$_{-0.02}^{+0.02}$	& 2.94$_{-0.04}^{+0.04}$	& 2.66$_{-0.04}^{+0.04}$ \\ 
A$_{PL}$ ($\times10^{-4}$)	& 10.5$_{-0.1}^{+0.2}$& 7.6$_{-0.1}^{+0.1}$& 5.2$_{-0.1}^{+0.1}$	& 4.6$_{-0.1}^{-0.1}$ \\
kT (eV)			& 108$_{-1}^{+1}$     		& 122$_{-1}^{+1}$& 96$_{-4}^{+3}$ 	& 109$_{-1}^{+1}$ \\
A$_{BB}$ ($\times10^{-5}$) & 7.2$_{-0.2}^{+0.2}$	& 8.4$_{-0.2}^{+0.2}$	& 4.0$_{-0.2}^{+0.2}$	& 4.4$_{-0.2}^{+0.2}$ \\
E$_{\kalpha}$ (\kev)	& 6.65$_{-0.05}^{+0.05}$	&	--	& 7.43$_{-0.26}^{+0.27}$	&   --	\\
Index ($\beta$)  $^a$ 	& 3.65$_{-0.11}^{+0.08}$	&	--	& 5.1$_{-0.3}^{+0.3}$      	&   --	\\
R$_{in}$ (R$_g$) $^a$  	& \textless{} 1.7		&       --	& 1.77$_{-0.11}^{+0.08}$		&   --  \\
R$_{out}$ (R$_g$) $^a$	& 400*     			&       --	& 400*  			&   --  \\
Inclination (degree) $^a$& 30*				&       --	& 30*				&   --  \\
$f_{K_{\alpha}}$  $^b$ ($\times10^{-5}$)& 2.28$_{-0.21}^{+0.27}$& --	& 1.4$_{-0.1}^{+0.2}$		& -- 	\\
E$_{\lalpha}$ (\kev)	& 0.92$_{-0.01}^{+0.01}$ 	& 	--	& 1.1$_{-0.1}^{+0.1}$		&   --  \\
$f_{L_{\alpha}}$ $^b$ ($\times10^{-4}$)	& 4.86$_{-0.25}^{+0.31}$& --	& 3.66$_{-0.51}^{+0.68}$	&-- 	\\\hline
$\chi^2 / dof$  	& 489 / 342			& 314 / 284	& 351 / 350			& 238 / 243 \\\hline
\end{tabular}\label{spec_res}
\end{center}
{\footnotesize Notes: ($a$) lines parameters that are tied between Fe L and Fe K lines and ($b$) indicate line flux in units of Photons cm$^{-2}$ s$^{-1}$.}
\end{table*}

\section{DISCUSSION AND CONCLUSION}
We performed time resolved spectroscopy of NLS1 galaxies \hhh{} and \iras{} using $\sim$ 100 $ks$ long \xmm{} observations. NLS1 galaxies are generally characterized by steep power law spectrum and rapid variability. 
We found that both these AGN show strong short--term variations within the observations. Figure~\ref{lc_variations} shows the significant variability observed in the soft band (0.3--1 \kev{}), the hard band (1.5--5 \kev{}) and in the hardness ratio. The variability in hardness ratio suggests the spectral variability within the observation. These spectral variations motivated us to study the variability of spectral components and the artifacts which could be introduced by such variability in the average spectrum. In this study we performed time resolved spectroscopy by generating spectra from multiple small segments. 
In Fig.~\ref{spec_examples} we show unfolded spectra of two typical segments fitted with a $power$ $law$ model (with $\Gamma$ fixed at 2) to show the short--term spectral variations in both the sources. This figure reveals the significant spectral variability within the observation.
We find that each segment spectra can be easily modelled using two component model (PL+BB) modified by the Galactic absorption. As expected, our spectral result exhibit that all the spectral parameters were variable and are plotted in Fig.~\ref{spec_variations}. Variability of $\Gamma_{PL}$ and kT$_{BB}$ indicates that the observed variability is not just due to the flux variations but the spectral shape is also changing.\\
The best fit model of each spectra were used to simulate the data which were co--added to get the combined simulated spectra. These combined simulated spectra were later compared with the actual time averaged spectra. The result of this spectral fitting is tabulated in Table~\ref{spec_res}. The spectral parameters show variations between the simulated and average spectra, which is solely because the GTI of simulated spectrum is a subset of total exposure of the time averaged spectrum. The extra time in the average spectra caused the variation in the spectral parameters seen in Table~\ref{spec_res}. We find that the time averaged spectra showed significant deviations and required additional line components, therefore we added two $Laor$ lines to the model. 
In simulated spectra we do not see line at 6.4 \kev{} as our initial model did not have the line model, however, no deviation was seen near 0.9 \kev{}. This suggest that the line we see in the time averaged spectra is not the artifact of the variation of spectral components. 
In fact, no deviation near 0.9 \kev{}, this is an independent way of proving that the line indeed is a genuine feature. Even if we did not find line feature but a positive deviation around 0.9 \kev{} would have certainly lowered the overabundance required in these objects and would have improved the current reflection models.

\section{ACKNOWLEDGEMENT}
We thank the anonymous referee for his/her constructive comments. PKP acknowledges financial support from CSIR, New Delhi. SKJ acknowledges financial support from DST, New Delhi through the INSPIRE Scheme. This research has made use of the observations from \xmm{} telescope and analysis is carried out using the software provided by Science Analysis System (SAS) and the \ftool{} provided by the High Energy Astrophysics Science Archive Research Center (HEASARC) software package. First three figures were generated using \emph{veusz} plotting tool.\\

\def\aj{AJ} \def\actaa{Acta Astron.}  \def\araa{ARA\&A} \def\apj{ApJ}
\def\apjl{ApJ} \def\apjs{ApJS} \def\ao{Appl.~Opt.}  \def\apss{Ap\&SS}
\def\aap{A\&A} \def\aapr{A\&A~Rev.}  \def\aaps{A\&AS} \def\azh{AZh}
\def\baas{BAAS} \def\bac{Bull. astr. Inst. Czechosl.}
\def\caa{Chinese Astron. Astrophys.}  \def\cjaa{Chinese
  J. Astron. Astrophys.}  \def\icarus{Icarus} \def\jcap{J. Cosmology
  Astropart. Phys.}  \def\jrasc{JRASC} \def\mnras{MNRAS}
\def\memras{MmRAS} \def\na{New A} \def\nar{New A Rev.}
\def\pasa{PASA} \def\pra{Phys.~Rev.~A} \def\prb{Phys.~Rev.~B}
\def\prc{Phys.~Rev.~C} \def\prd{Phys.~Rev.~D} \def\pre{Phys.~Rev.~E}
\def\prl{Phys.~Rev.~Lett.}  \def\pasp{PASP} \def\pasj{PASJ}
\def\qjras{QJRAS} \def\rmxaa{Rev. Mexicana Astron. Astrofis.}
\def\skytel{S\&T} \def\solphys{Sol.~Phys.}  \def\sovast{Soviet~Ast.}
\def\ssr{Space~Sci.~Rev.}  \def\zap{ZAp} \def\nat{Nature}
\def\iaucirc{IAU~Circ.}  \def\aplett{Astrophys.~Lett.}
\def\apspr{Astrophys.~Space~Phys.~Res.}
\def\bain{Bull.~Astron.~Inst.~Netherlands}
\def\fcp{Fund.~Cosmic~Phys.}  \def\gca{Geochim.~Cosmochim.~Acta}
\def\grl{Geophys.~Res.~Lett.}  \def\jcp{J.~Chem.~Phys.}
\def\jgr{J.~Geophys.~Res.}
\def\jqsrt{J.~Quant.~Spec.~Radiat.~Transf.}
\def\memsai{Mem.~Soc.~Astron.~Italiana} \def\nphysa{Nucl.~Phys.~A}
\def\physrep{Phys.~Rep.}  \def\physscr{Phys.~Scr}
\def\planss{Planet.~Space~Sci.}  \def\procspie{Proc.~SPIE}
\let\astap=\aap \let\apjlett=\apjl \let\apjsupp=\apjs \let\applopt=\ao

\bibliographystyle{raa} \bibliography{ms0100}
\label{lastpage}

\end{document}